# Electron scattering cross section for light nuclei within the unified electroweak theory

Vo Minh Truong[1,2,*] and Nguyen Quang Hung[1,2]

[1]Institute of Fundamental and Applied Sciences, Duy Tan University, Ho Chi Minh City 70000, Vietnam
[2]Faculty of Natural Sciences, Duy Tan University, Danang City 50000, Vietnam

We employ the multipole expansion within the unified electroweak theory to develop a complete calculation method of the electron scattering cross section for light nuclei. The specific calculations for $^{6,7}$Li and $^{7}$Be nuclei indicate that the conventional impulse approximation can be applied to the electron-nucleus scattering only when the incident electron energy does not exceed twice the nucleon mass. In addition, the quasi-elastic scattering cross section of $^{7}$Li in the excitation from its ground state to the nearest excited state and that of elastic scattering in the ground state are independently treated, whereas they have not been separately measured in experiments. The obtained scattering cross sections corresponding to an appropriate adjustment of the harmonic oscillator parameter along with the V-A structure interpret reasonably well the available experimental data at MeV energies and provide predictive information at GeV energies. This gives new perspectives in studying the structure of nuclei and the weak interactions via electron or lepton scattering at high (GeV) energies, considering nuclei rather than quarks.

## I. INTRODUCTION

Electron scattering is an effective tool for studying the structure of nuclei, especially at low energies when the weak interactions are negligible [1-5]. A primary quantity in theoretical studies is the scattering cross section, determined by the square of the scattering amplitude. Due to the unification of the electromagnetic and weak interactions at high energies [6] together with the development of electron accelerators [7,8], which is sufficient to test such a unification, the calculation of electron-nucleus scattering cross section using the unified electroweak theory has been feasible. One of the feasibilities is to expand the transition currents inside the nucleus into the multipole components, each with a certain angular momentum, the so-called multipole form factor [2,3,9].

Weigert and Rose [3] derived the first complete multipole expansion of the electron-nucleus scattering cross section at low energies, in which the nuclear recoil was ignored and the electron polarization was calculated within the non-relativistic theory. In their study, a general expression for the scattering cross section was proposed but without a specific calculation for the multipole form factors. The latter were treated as momentum-transfer-dependent parameters, whose values were inferred by comparing the theoretical cross sections with the experimental data. Meanwhile, Willey [2] performed the multipole expansion for the electromagnetic scattering of unpolarized electrons and unoriented nuclei, in which the conditions for applying the approximations as well as the explicit formulae for calculating the multipole form factors were introduced. However, the angular matrix elements of the product $\mathbf{Y}_L^{L'}.\mathbf{l}$ ($\mathbf{Y}_L^{L'}$ are the spherical harmonics for the multipoles $L$ and $L'$, $\mathbf{l}$ is the angular momentum) for multiparticle cases were not specifically mentioned in this study, while other formulae were calculated using the nuclear shell model, fractional parentage coefficients, and the angular momentum theory. The two studies [2,3] have provided a complete calculation for the electron-nucleus scattering cross section at low energies. Application of Willey's formulae to calculate the multipole form factors in electron-nucleus scattering is convenient since

***Corresponding author**: Vo Minh Truong; Institute of Fundamental and Applied Sciences, Duy Tan University, Ho Chi Minh City 70000, Vietnam; Email: vominhtruong@duytan.edu.vn

no additional assumption is required, while the fractional parentage coefficients are mostly available for use [10-12].

To determine the scattering cross section within the unified electroweak theory, the multiparticle matrix elements must also be calculated. Moreover, in addition to the conventional matrix elements of the electromagnetic multipoles, those of the vector and axial multipoles are required [13-15]. Donnelly *et al.* [16-18] implemented separate calculations for the electromagnetic and weak transition currents without arriving at a complete multipole expansion for the electron scattering cross section. The total cross section can be treated as a sum of the cross sections due to the electromagnetic and weak interactions. Furthermore, the multiparticle matrix elements were transformed into the single-particle ones using the one-body density matrix elements or spectral function. Those transformations made the calculations to be approximate. In later studies by Luong [14,15], the weak and electromagnetic interactions were equally treated, while the multipole expansions for the unified electroweak current densities were also implemented, thereby arriving at the scattering cross section in terms of the multipole form factors. As a result, besides the cross sections of the pure electromagnetic and weak interactions, there was an interferential term between them. In this approach, the scattering of oriented nuclei by leptons was also investigated. However, these studies still employed the impulse approximation previously developed by Kerimov *et al.* [13]. Accordingly, the convection currents were neglected and the axial form factors were determined via the electromagnetic ones. Nevertheless, it was pointed out that such an impulse approximation was only suitable for nucleon-nucleon scattering in free space, which is limited to the regions of intermediate energies and small scattering angles [19]. In addition, the calculations of multiparticle matrix elements using the particle density matrix and spectral function are convenient only for nuclei with a few nucleons. For nuclei with more nucleons, the problem becomes more complicated, and additional assumptions are required.

In our recent work [20], a calculation of the electron-nucleus scattering cross section for polarized electrons and unoriented nuclei has been developed using the unified electroweak theory. The matrix elements of all multipole operators have been directly calculated based on the many-particle shell model and fractional parentage coefficients. The numerical calculations for the elastic and quasi-elastic scattering cross sections of a stable $^{7}$Li nucleus using the Weinberg-Salam model and the given harmonic oscillator parameter have indicated the non-negligible contribution of weak interactions when the incident electron energy is above about 1 GeV.

This study extends our previous work to be able to apply for more light nuclei. For illustration, we choose two stable $^{6,7}$Li and one unstable $^{7}$Be nuclei, whose structure can be well described by the many-particle shell model. Those nuclei are selected because the differences by one nucleon in their outer shell might lead to some features that can be found when analyzing the corresponding electron scattering cross sections. In addition, the harmonic oscillator parameters are deduced by comparing the theoretical cross sections with the experimental data at a specific energy and angle [4,5]. Furthermore, because the center of mass and the self-consistent potential field are not fixed during the scattering, the calculated cross sections will be compensated by a factor in the form of a Gaussian function.

## II. FORMALISM

Multipole expansion for the scattering cross section has been presented in detail in our previous study [20], so we present in this section only its final forms. In calculations, we set $\hbar = c = 1$ for simplicity. The cyclic coordinate system is also used to derive the expressions of multipole expansion for the electroweak current densities and corresponding cross sections. For the weak currents, we take into account the contribution of neutral weak currents only. In addition, the multipole operators are formulated within the framework of the many-particle shell model, while the reduced matrix elements are deduced by using the one-particle fractional parentage coefficients.

### 2.1. Electron scattering cross section

The electron-nucleus scattering cross section can be calculated by using the Born approximation, in which the incoming electron is supposed to exchange a photon $\gamma$ and an intermediate neutral boson $Z^0$ with the target nucleus in the scattering process. In addition, the target nucleus must be light to satisfy the inequality $Z\alpha < 1$ with $\alpha$ being the electromagnetic coupling constant. The scattering cross section can be written in the form as [20]

$$\sigma = \frac{4m_e^2\varepsilon'}{f\varepsilon}\overline{\sum_{if}}|M_{fi}|^2 = \frac{\varepsilon'}{4f\varepsilon Q^4}(R_F + R_{FZ} + R_Z), \quad (1)$$

where $M_{fi}$ is the scattering amplitude, whose explicit form is given in Ref. [20]; $Q = K - K' = (\omega, \mathbf{q})$ is the momentum transfer with $K = (\varepsilon, \mathbf{k})$ and $K' = (\varepsilon', \mathbf{k}')$ being, respectively, the electron momenta before and after the scattering; $m_e$ is the electron mass; and $f$ is the nuclear recoil factor, which is the same as that in Ref. [21]. The notation $\overline{\sum_{if}}$ denotes the average over the initial spin states and the summation over the final states, performing for both electron and nucleus. Three cross section terms in Eq. (1) are the contraction products of an electron or lepton tensor and a nucleus or hadron tensor, whose exact forms can be obtained by expanding the summations over electron polarization states, namely [20]

$$R_F = 4\alpha^2[(1+\xi\xi')A_1 + (\xi+\xi')A_2], \quad (2a)$$

$$R_{FZ} = 8\alpha\lambda\{[g_V(1+\xi\xi') + g_A(\xi+\xi')]B_1 + [g_V(\xi+\xi') + g_A(1+\xi\xi')]B_2\}, \quad (2b)$$

$$R_Z = 4\lambda^2\{[(g_V^2+g_A^2)(1+\xi\xi') + 2g_V g_A(\xi+\xi')]C_1 + [(g_V^2+g_A^2)(\xi+\xi') + 2g_V g_A(1+\xi\xi')]C_2\}. \quad (2c)$$

In Eqs. (2a)-(2c), $g_V = -1/2 + 2x_W$, $g_A = -1/2$, and $x_W \equiv \sin^2\theta_W$ are the parameters of the weak interaction given in the Weinberg-Salam model with $\theta_W$ being the Weinberg angle. $\xi$ and $\xi'$ are, respectively the electron polarizations before and after the scattering, whose values are equal to ±1. $\lambda = -G_F m_Z^2 Q^2/[2\sqrt{2}\pi(m_Z^2 - Q^2)]$ with $G_F = g^2/4\sqrt{2}m_Z^2\cos^2\theta_W$ ($g$ is the weak coupling constant and $m_Z$ is the $Z^0$ boson mass) is the effective weak interaction constant. The coefficients $A_1$, $A_2$, $B_1$, $B_2$, $C_1$, and $C_2$, which depend on the products of two transition currents, are calculated by expanding them into the multipole form factors. Their explicit forms can be found in Ref. [20], so we do not repeat them here.

The expressions (1) and (2a)-(2c) provide a complete expansion for the electron-nucleus scattering cross section, in which the contribution of neutral weak currents is taken into account.

### 2.2. Multipole operators

To determine the scattering cross section, we need to compute the transition amplitudes of the multipole components between the initial and final states [20]. In addition, the nucleon form factors should be included in the expressions of the current densities to describe the nucleon size. At high energies, these form factors are parameterized in terms of dipoles, which depend on the momentum transfer. It has been known so far that the nucleon form factors include the electric $G_E$, magnetic $G_M$, axial $G_A$, induced pseudoscalar $G_P$, induced pseudotensor $G_T$, and gravitational components [22-29]. While the electromagnetic form factors are well-described by experimental data and theoretical calculations, many problems relating to the remaining form factors have not yet been clarified. In principle, $G_T$ can be presented in the charge-changing scattering processes. However, it disappears when combining the charge conjugation invariance and isospin symmetry. In addition, Refs. [25,26] have pointed out that $G_T$ is small and can be ignored. Meanwhile, $G_P$ appears only in the capture of the radiation and ordinary muons as well as the generation of charged pions. $G_P$ is also negligible according to the calculations within the chiral perturbation theory.

In the present work, the nuclear current densities are calculated by using the electroweak theory and the many-particle shell model. Those current densities can be found in Ref. [20], e.g., Eqs. (20a)-(20c), except that the axial currents are given with a minus sign for consistency to the V-A structure of the neutral weak currents. Thus, the explicit forms for the electromagnetic multipole operators are now given as [20]

$$\hat{F}^{C}_{Lm}(q)=i^{L}\sum_{a}X^{F}_{a}\,j_{L}(qr_{a})Y_{Lm}(\underline{\mathbf{r}}_{a})\,, \qquad (3a)$$

$$\hat{F}^{E}_{Lm}(q)=\frac{i^{L}q}{2m_{N}}\sum_{a}\Big\{\frac{X^{F}_{a}}{\sqrt{L(L+1)}}[r_{a}\,j_{L}(qr_{a})d_{a}$$

$$+d_{a}r_{a}\,j_{L}(qr_{a})]Y_{Lm}(\underline{\mathbf{r}}_{a})+Y^{F}_{a}\,j_{L}(qr_{a})\mathbf{Y}^{L}_{Lm}(\underline{\mathbf{r}}_{a}).\boldsymbol{\sigma}_{a}\Big\}\,, \quad (3b)$$

$$\hat{F}^{M}_{Lm}(q)=\frac{i^{L+1}q}{2m_{N}[L]}\sum_{a}[2X^{F}_{a}\mathbf{A}^{L}_{a}.\mathbf{l}_{a}+Y^{F}_{a}\mathbf{B}^{L}_{a}.\boldsymbol{\sigma}_{a}]\,, \quad (3c)$$

while the axial multipole operators read

$$\hat{A}^{C}_{Lm}(q)=\frac{i^{L+1}}{m_{N}}\sum_{a}Y^{A}_{a}\,j_{L}(qr_{a})Y_{Lm}(\underline{\mathbf{r}}_{a})\nabla_{a}.\boldsymbol{\sigma}_{a}\,, \qquad (4a)$$

$$\hat{A}^{E}_{Lm}(q)=-\frac{i^{L+1}}{[L]}\sum_{a}Y^{A}_{a}\mathbf{B}^{L}_{a}.\boldsymbol{\sigma}_{a}\,, \qquad (4b)$$

$$\hat{A}^{M}_{Lm}(q)=-i^{L}\sum_{a}Y^{A}_{a}\,j_{L}(qr_{a})\mathbf{Y}^{L}_{Lm}(\underline{\mathbf{r}}_{a}).\boldsymbol{\sigma}_{a}\,. \qquad (4c)$$

The vector multipoles can be deduced directly from the electromagnetic ones. We shortly use $[L]\equiv\sqrt{2L+1}$, $d_{a}\equiv d/dr_{a}$, $\mathbf{A}^{L}_{a}\equiv\mathbf{A}^{L}(\mathbf{r}_{a})$, $\mathbf{B}^{L}_{a}\equiv\mathbf{B}^{L}(\mathbf{r}_{a})$ with $\mathbf{A}^{L}(\mathbf{r})$ and $\mathbf{B}^{L}(\mathbf{r})$ being given in Ref. [20], i.e., Eqs. (23a)-(23b). Here, $j_{L}(qr_{a})$, $Y_{Lm}(\underline{\mathbf{r}}_{a})$, and $\mathbf{Y}^{L}_{Lm}(\underline{\mathbf{r}}_{a})$ ($\underline{\mathbf{r}}$ is the unit vector along the direction of $\mathbf{r}$) are the spherical Bessel, spherical function, and spherical vector, respectively.

It can be seen that if the convection currents are neglected as in the impulse approximation, the electric multipoles of the axial current will be proportional to the magnetic multipoles of the electromagnetic current. Meanwhile, the magnetic multipoles of the axial current will also be proportional to the electric multipoles of the electromagnetic one. To determine the multipole form factors presented in $A_1$, $A_2$, $B_1$, $B_2$, $C_1$, and $C_2$ and in Eqs. (2a)-(2c), the reduced matrix elements of the operators (3a)-(3c) and (4a)-(4c) are needed to be calculated. Since the single-particle matrix elements were provided in Refs. [2,18], we should calculate the two-particle matrix elements as a separate case as well as the multiparticle matrix elements as a general one.

### 2.3. Reduced matrix elements

In the many-particle shell model, nucleons in the inner shell form a spherical core of spin 0 and link with the valence nucleons. Various studies have pointed out that the energy spectrum and many characteristics of nuclei that obey the shell model can be elucidated when considering nucleons in the outer shells only [1,2,4]. Therefore, only valence nucleons in the unfilled shell are often taken into account in calculations. In the present study, the two-particle matrix elements are treated as a special case because they can be directly converted to the single-particle ones. The total anti-symmetric wavefunction is written as a product of the orbital, spin and isospin wavefunctions since nucleons are supposed to move in a spherically symmetric field. To derive the reduced matrix elements, we use the notations $|i\rangle\equiv|n_{i}L_{i}S_{i}T_{i}M_{T_{i}}J_{i}\rangle$ and $|f\rangle\equiv|n_{f}L_{f}S_{f}T_{f}M_{T_{f}}J_{f}\rangle$ for the initial and final states in *LS* coupling, and the braces $\{\ldots\}$ for the 6*j* or 9*j* coefficients. The spherical Bessel function acts on the radial wavefunction, while the spherical vector and the angular momentum act on the orbital wavefunction. The operator $\boldsymbol{\sigma}$ acts on the spin wavefunction and the operators $X^{S}_{a}$ and $Y^{S}_{a}$ act only on the isospin wavefunction.

The matrix elements of an identical two-particle system in the $nl^{2}$ state have the final forms as

$$\langle f\|\sum_{a}X^{F}_{a}\Im_{L}(a)\|i\rangle=2\delta_{S_{i}S_{f}}\,\delta_{M_{T_{i}}M_{T_{f}}}(-1)^{L_{i}+S_{i}+J_{i}+L_{f}+l_{i}+l_{f}}[L_{i}][L_{f}]$$

$$\times[J_{i}][J_{f}]\begin{Bmatrix}L_{i} & S_{i} & J_{i}\\ J_{f} & L & L_{f}\end{Bmatrix}\begin{Bmatrix}l_{i} & l_{i} & L_{i}\\ L_{f} & L & l_{f}\end{Bmatrix}\langle n_{f}l_{f}\|\Im_{L}\|n_{i}l_{i}\rangle Q^{F}, \quad (5a)$$

$$\langle f\|\sum_{a}Y^{F}_{a}\,j_{L'}(qr_{a})\mathbf{Y}^{L'}_{L}(\underline{\mathbf{r}}_{a}).\boldsymbol{\sigma}_{a}\|i\rangle=2\sqrt{6}\delta_{M_{T_{i}}M_{T_{f}}}(-1)^{L_{i}+S_{i}+L'+l_{i}+l_{f}}[L_{i}][L_{f}]$$

$$\times[L][S_{i}][S_{f}][J_{i}][J_{f}]\begin{Bmatrix}L_{i} & S_{i} & J_{i}\\ L' & 1 & L\\ L_{f} & S_{f} & J_{f}\end{Bmatrix}\begin{Bmatrix}l_{i} & l_{i} & L_{i}\\ L_{f} & L' & l_{f}\end{Bmatrix}$$

$$\times\begin{Bmatrix}1/2 & 1/2 & S_{i}\\ S_{f} & 1 & 1/2\end{Bmatrix}\langle n_{f}l_{f}\|j_{L'}(qr)\mathbf{Y}^{L'}(\underline{\mathbf{r}})\|n_{i}l_{i}\rangle Q^{F}, \qquad (5b)$$

$$\langle f\|\sum_{a}X^{F}_{a}\,j_{L'}(qr_{a})\mathbf{Y}^{L'}_{L}(\underline{\mathbf{r}}_{a})\mathbf{l}_{a}\|i\rangle=2\delta_{S_{i}S_{f}}\,\delta_{M_{T_{i}}M_{T_{f}}}(-1)^{L_{i}+S_{i}+J_{i}+L_{f}+L}[L]$$

$$\times[L_{i}][L_{f}][J_{i}][J_{f}]\begin{Bmatrix}L_{i} & S_{i} & J_{i}\\ J_{f} & L & L_{f}\end{Bmatrix}\begin{Bmatrix}l_{i} & l_{i} & L_{i}\\ L_{f} & L & l_{f}\end{Bmatrix}\begin{Bmatrix}L' & L & 1\\ l_{i} & l_{i} & l_{f}\end{Bmatrix}$$

$$\times\langle l_{i}\|\mathbf{l}\|l_{i}\rangle\langle n_{f}l_{f}\|j_{L'}(qr)\mathbf{Y}^{L'}(\underline{\mathbf{r}})\|n_{i}l_{i}\rangle Q^{F}, \qquad (5c)$$

$$\langle i\|\sum_a Y_a^A j_{L'}(qr_a)\mathbf{Y}_L^{L'}(\underline{\mathbf{r}}_a).\boldsymbol{\sigma}_a\|f\rangle = 2\sqrt{6}\delta_{M_{T_i}M_{T_f}}(-1)^{L_f+S_f+L'+l_i+l_f}[L]$$

$$\times[L_i][L_f][S_i][S_f][J_i][J_f]\begin{Bmatrix} L_f & S_f & J_f \\ L' & 1 & L \\ L_i & S_i & J_i \end{Bmatrix}\begin{Bmatrix} l_f & l_f & L_f \\ L_i & L' & l_i \end{Bmatrix}$$

$$\times\begin{Bmatrix} 1/2 & 1/2 & S_f \\ S_i & 1 & 1/2 \end{Bmatrix}\langle n_i l_i\| j_{L'}(qr)\mathbf{Y}^{L'}(\underline{\mathbf{r}})\|n_f l_f\rangle Q^A, \qquad (5d)$$

where the matrix elements of the isospin wavefunctions are given as

$$Q^F = \delta_{T_iT_f} G^s_{E,M} + (-1)^{T_i}\sqrt{6}[T_f]C^{T_fM_{T_f}}_{T_iM_{T_i}10} G^v_{E,M}, \qquad (6a)$$

$$Q^A = \delta_{T_iT_f}\beta_A^{(0)} G^s_A + (-1)^{T_i}\sqrt{6}[T_f]C^{T_fM_{T_f}}_{T_iM_{T_i}10}\beta_A^{(1)} G^v_A. \qquad (6b)$$

Here $\Im_L$ stands for the operators acting on the orbital coordinates. There is a factor of two that appeared in Eqs. (5a)-(5d) because the associated summations are taken over the number of particles. The above formulae enable us to calculate all the matrix elements of the multipole operators when considering two valence nucleons in the outer shell.

In principle, the multiparticle matrix elements can be calculated by transforming them into single-particle ones in terms of a linear combination using particle-density matrix elements, spectral function, or fractional parentage coefficients. In Ref. [20], we have used the many-particle shell model and fractional parentage coefficients to derive those multiparticle matrix elements. The one-particle fractional parentage coefficients [2,10] were employed to deduce the general formulae as all the multipole operators depend only on the single-particle variables. However, these formulae should be modified and extended to be applicable to a variety of nuclei, including both odd and even ones. As a result, the final matrix elements of an identical $A'$ particle system in the $nl^{A'}$ state can be written in the more symmetric forms as

$$\left\langle f\left\|\sum_a X_a^F \Im_L(a)\right\|i\right\rangle = \delta_{S_iS_f}\delta_{M_{T_i}M_{T_f}} A'\sum_P (-1)^{L_P+S_i+J_i+l_i}[L_i][L_f]$$

$$\times[J_i][J_f]\langle\psi_i\{|\psi_P\rangle\langle\psi_f\{|\psi_P\rangle\begin{Bmatrix} l_i & L_P & L_i \\ L_f & L & l_f \end{Bmatrix}$$

$$\times\begin{Bmatrix} L_i & S_i & J_i \\ J_f & L & L_f \end{Bmatrix}\langle n_f l_f\|\Im_L\|n_i l_i\rangle Q_P^F, \qquad (7a)$$

$$\left\langle f\left\|\sum_a Y_a^F j_{L'}(qr_a)\mathbf{Y}_L^{L'}(\underline{\mathbf{r}}_a).\boldsymbol{\sigma}_a\right\|i\right\rangle = \delta_{M_{T_i}M_{T_f}} A'\sum_P (-1)^{L_P+S_P+L'+L_f+S_f+l_i+3/2}$$

$$\times\sqrt{6}[L][L_i][L_f][S_i][S_f][J_i][J_f]\langle\psi_i\{|\psi_P\rangle\langle\psi_f\{|\psi_P\rangle\begin{Bmatrix} L_i & S_i & J_i \\ L' & 1 & L \\ L_f & S_f & J_f \end{Bmatrix}$$

$$\times\begin{Bmatrix} l_i & L_P & L_i \\ L_f & L' & l_f \end{Bmatrix}\begin{Bmatrix} 1/2 & S_P & S_i \\ S_f & 1 & 1/2 \end{Bmatrix}\langle n_f l_f\| j_{L'}(qr)\mathbf{Y}^{L'}(\underline{\mathbf{r}})\|n_i l_i\rangle Q_P^F, \qquad (7b)$$

$$\left\langle f\left\|\sum_a X_a^F j_{L'}(qr_a)\mathbf{Y}_L^{L'}(\underline{\mathbf{r}}_a).\mathbf{l}_a\right\|i\right\rangle = \delta_{S_iS_f}\delta_{M_{T_i}M_{T_f}} A'\sum_P (-1)^{L_P+L+S_i+J_i+l_f}$$

$$\times[L][L_i][L_f][J_i][J_f]\langle\psi_i\{|\psi_P\rangle\langle\psi_f\{|\psi_P\rangle\begin{Bmatrix} L_i & S_i & J_i \\ J_f & L & L_f \end{Bmatrix}\begin{Bmatrix} l_i & L_P & L_i \\ L_f & L & l_f \end{Bmatrix}$$

$$\times\begin{Bmatrix} L' & L & 1 \\ l_i & l_i & l_f \end{Bmatrix}\langle l_i\|\mathbf{l}\|l_i\rangle\langle n_f l_f\| j_{L'}(qr)\mathbf{Y}^{L'}(\underline{\mathbf{r}})\|n_i l_i\rangle Q_P^F, \qquad (7c)$$

$$\left\langle i\left\|\sum_a Y_a^A j_{L'}(qr_a)\mathbf{Y}_L^{L'}(\underline{\mathbf{r}}_a).\boldsymbol{\sigma}_a\right\|f\right\rangle = \delta_{M_{T_i}M_{T_f}} A'\sum_P (-1)^{L_P+S_P+L'+L_i+S_i+l_f+3/2}$$

$$\times\sqrt{6}[L][L_i][L_f][S_i][S_f][J_i][J_f]\langle\psi_i\{|\psi_P\rangle\langle\psi_f\{|\psi_P\rangle\begin{Bmatrix} L_f & S_f & J_f \\ L' & 1 & L \\ L_i & S_i & J_i \end{Bmatrix}$$

$$\times\begin{Bmatrix} l_f & L_P & L_f \\ L_i & L' & l_i \end{Bmatrix}\begin{Bmatrix} 1/2 & S_P & S_f \\ S_i & 1 & 1/2 \end{Bmatrix}\langle n_i l_i\| j_{L'}(qr)\mathbf{Y}^{L'}(\underline{\mathbf{r}})\|n_f l_f\rangle Q_P^A, \qquad (7d)$$

where

$$Q_P^F = \delta_{T_iT_f} G^s_{E,M} + (-1)^{T_P+T_f+3/2}\sqrt{6}[T_i]C^{T_fM_{T_f}}_{T_iM_{T_i}10}\begin{Bmatrix} 1/2 & 1 & 1/2 \\ T_f & T_P & T_i \end{Bmatrix} G^v_{E,M}, \qquad (8a)$$

$$Q_P^A = \delta_{T_iT_f}\beta_A^{(0)} G^s_A + (-1)^{T_P+T_f+3/2}\sqrt{6}[T_i]C^{T_fM_{T_f}}_{T_iM_{T_i}10}\begin{Bmatrix} 1/2 & 1 & 1/2 \\ T_f & T_P & T_i \end{Bmatrix}\beta_A^{(1)} G^v_A. \qquad (8b)$$

Here, all the summations are taken over the parent states $P = L_P, S_P, T_P$ and $\langle\psi\{|\psi_P\rangle$ denotes the one-particle fractional parentage coefficients. The isospin matrix elements $Q_P^{F,A} = Q_0^{F,A}$, $Q_{1/2}^{F,A}$, and $Q_1^{F,A}$ correspond to $T_P$ = 0, 1/2, and 1. For all light nuclei, $T_P$ = 0, 1 when the number of particles in the parent states is even, while $T_P$ = 1/2 when it is an odd-number case.

The electromagnetic and axial form factors for two or more nucleons can be completely calculated, thanks to the formulae above. In addition, the vector form factors can be determined by replacing $X_a^F, Y_a^F \to X_a^V, Y_a^V$. We note that the Coulomb axial form factors are computed by integrating in parts (4a) and then applying (5d) and (7d).

## III. RESULTS

The formulae (5a)-(5d) and (7a)-(7d) allow us to determine multipole form factors of light nuclei with arbitrary spins. Here we choose the stable nucleus $^6$Li with spin 1 to calculate for the separate case. The stable $^7$Li and unstable $^7$Be nuclei with the same spin 3/2 are selected for the general case. The numerical calculation results for the case of unpolarized electrons will be analyzed within the Weinberg-Salam model, in which the cross section (2) is multiplied by a factor of two as it was averaged over the initial electron states. The nucleon form factors are given by $G_E^p = -G_E^n/1.91\eta = G_M^p/2.79 = -G_M^n/1.91 = G_D$, $G_D = 1/(1+q^2/0.71\text{GeV}^2)^2$, $\eta = -Q^2/4m_N^2$, $G_A = G_A(0)/\,(1+q^2/m_A^2)^2$, where $G_A(0) = 1.27$ and $m_A = 1.026$ GeV [23-25]. The radial integrals in all cases are $J_0 = (1-2z/3)e^{-z}$, $J_2 = 2ze^{-z}/3$, $z = q^2/4\beta$, where $\beta$ being the harmonic oscillator parameter [2].

Tassie and Barker [30] pointed out that if the shell model well describes the nucleus and nucleons are supposed to move in a harmonic oscillator potential, the effect of the center of mass degree of freedom can be factorized, in which the matrix elements of the multipole operators need to be multiplied by a simple Gaussian function $e^{z/A}$ ($A$ is the nuclear mass number). Therefore, we multiply the calculated scattering cross sections by a compensating factor $e^{q^2/2\beta A}$. In addition, the experimental data in Refs. [4,5] indicated that at each incident electron energy, the experimental cross sections exhibited various errors when being measured at different angles. Thus, we shall determine each value of the harmonic oscillator parameter for each target nucleus by fitting the calculated cross sections to the experimental data having the smallest errors.

### 3.1. Electron elastic scattering on $^6$Li

We first consider the electron elastic scattering on the $^6$Li nucleus. The latter is in the ground state with two valence nucleons on the 1$p$ shell. This nucleus has a spin of 1, so according to the selection rule, the possible multipoles should be $0 \le L \le 2$. The multipole operators containing the factor $i^L$ in their expressions are thus even multipoles, while those with the factor $i^{L+1}$ are all odd multipoles. The multipoles $\hat{F}_{Lm}^E$, $\hat{V}_{Lm}^E$ and $\hat{A}_{Lm}^M$ are absent in elastic scattering due to the parity selection rule [9]. The reduced matrix elements of the operators (3a)-(3c) and (4a)-(4c) are calculated corresponding to the state indices, namely $n_i = n_f = 1, L_i = L_f = 0$, $l_i = l_f = 1, S_i = S_f = 1, J_i = J_f = 1, T_i = T_f = 0, M_{T_i} = M_{T_f} = 0$. The state of the two-nucleon system is $^{13}S$ with the convention $^{2T+1,2S+1}L$ according to the $LS$ coupling. The Clebsch-Gordan coefficients in (6a)-(6b) are equal to zero, so the isovector components in the expressions of the form factors are also equal to zero. The multipole form factors are obtained as

$$F_0^C(q) = \frac{\sqrt{3}}{\sqrt{\pi}} G_E^s J_0, \tag{9a}$$

$$F_2^C(q) = 0, \tag{9b}$$

$$F_1^M(q) = -\frac{q}{\sqrt{\pi} m_N} G_M^s J_0, \tag{9c}$$

$$A_1^C(q) = \frac{\sqrt{2}q}{\sqrt{\pi} m_N} \beta_A^{(0)} G_A^s J_0, \tag{9d}$$

$$A_1^E(q) = \frac{2}{\sqrt{\pi}} \beta_A^{(0)} G_A^s J_0. \tag{9e}$$

We do not write the vector form factors explicitly because they can be deduced from (9a)-(9c). All the nuclear form factors depend on those of nucleon along with the parameters of the unified electroweak theory. The radial integrals can be calculated in detail by using Watson's formula given in Ref. [2], noting that the isovector component is absent in the multipole form factors. Therefore, when applying the Weinberg-Salam model, all the axial form factors are equal to zero due to the parameter $\beta_A^{(0)} = 0$. This coincides with the general results for nuclei having the same numbers of protons and neutrons.

To achieve the numerical values for the scattering cross sections of $^6$Li, we need to determine first the harmonic oscillator parameter. This is done by adjusting $\beta$ so that the calculated cross section fits the experimental data which contain the smallest uncertainty, namely the data at 600 MeV and $32^0$ as shown in Table I of Ref. [5] ($\sigma_{\exp}$ = (1.48 ± 0.06) x $10^{-31}$ $cm^2/sr$). The obtained value is $\beta$ = 1.1526 $fm^{-2}$, at which the calculated cross section at the same energy (600 MeV) and scattering angle ($32^0$) is $\sigma_{\text{theor}}$ = 1.4815 x $10^{-31}$ $cm^2/sr$ (see Table 1). This value of $\beta$ is then used for calculating the numerical cross sections at other energies and scattering angles, whose results are presented in Tables 1 and 2. Results in Tables 1 and 2 reveal that below 1 GeV, the scattering cross sections are distributed at all angles, but the scattering processes occur only at small angles when the incident electron energy goes above 1 GeV. Especially, at hundreds of GeV and higher, the calculated cross sections differ from zero only when $\theta \simeq 0^0$, implying that incoming electrons almost go straightforward.

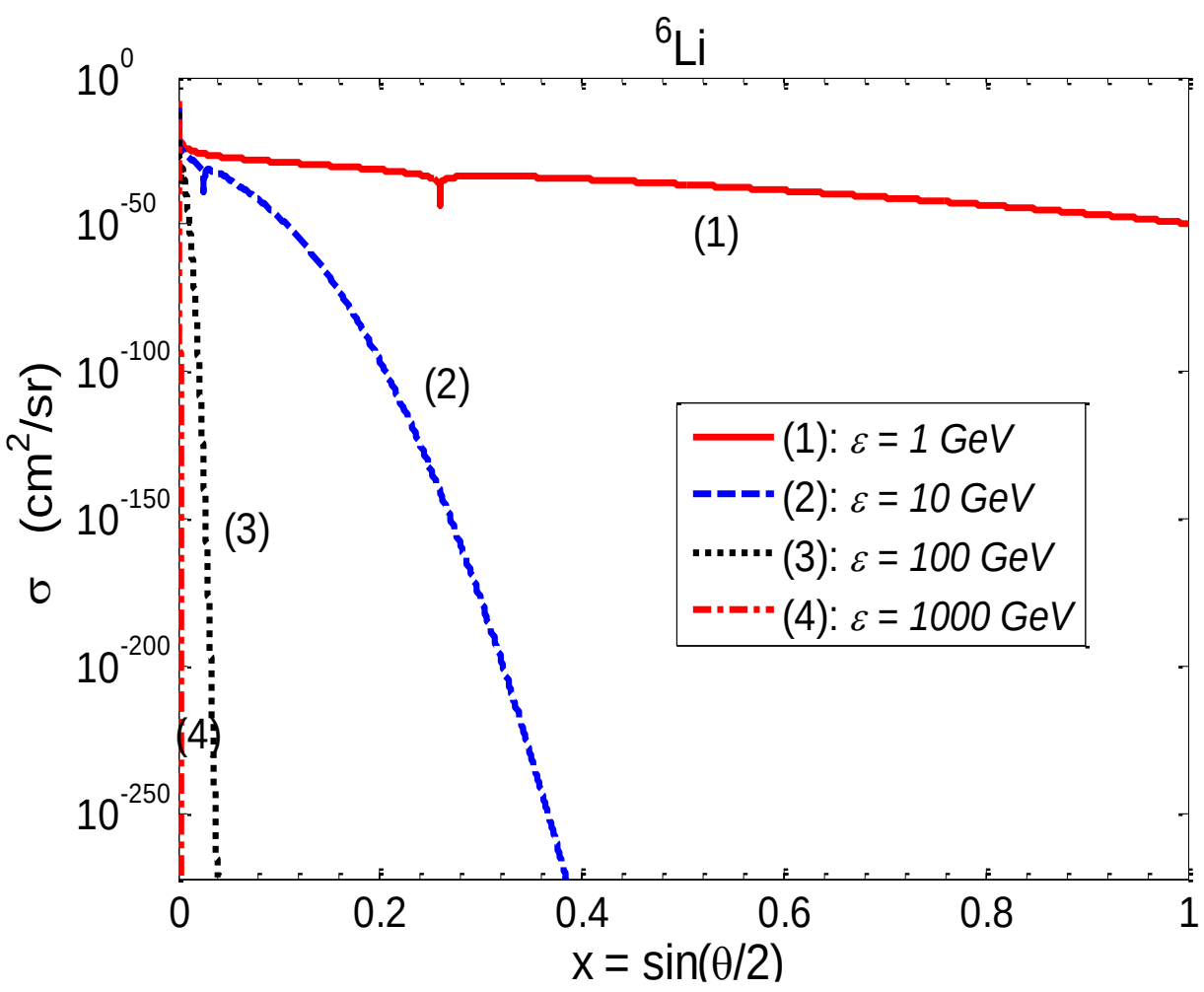

Fig. 1. Electron elastic scattering cross sections for $^6$Li obtained within the unified electroweak theory at 1, 10, 100, and 1000 GeV and different scattering angles.

Fig. 1 shows the plots of calculated scattering cross sections at above 1 GeV. One can see a general transformation rule that the cross section decreases with increasing both energy and scattering angle. In addition, the high-energy scattering processes tend to favor small angles because the cross sections diminish very quickly at large scattering angles. At 1 GeV, despite being still distributed at all angles, the cross section is much smaller than that obtained at the energies of MeV. The magnitude of the cross section rapidly decreases from the order of $10^{-17}$ at $0^0$ to $10^{-50}$ at $180^0$ as seen in Table 1. At 10 GeV, the cross section is nearly equal to zero, except at very small angles. The two curves (1) and (2) also have the extreme points commonly found in the electron elastic scattering [1]. At 100 and 1000 GeV, only forward scattering is possible, at which the cross sections have the magnitude of the orders $10^{-21}$ and $10^{-23}$ as listed in Table 2.

### 3.2. Electron elastic scattering on $^7$Li and $^7$Be

The calculations of elastic scattering cross section are then extended to the general case by considering two $^7$Li and $^7$Be mirror nuclei, whose ground-state spins are both equal to 3/2. These nuclei have similar energy spectra but with opposite isospin projections due to the difference in proton and neutron numbers. In this case, Eqs. (7a)-(7d) are used and three valance nucleons in the 1$p$ shell are considered. According to the selection rule, possible multipoles are $L = 0, 1, 2, 3$, and $\hat{F}^E_{Lm}$, $\hat{V}^E_{Lm}$ and $\hat{A}^M_{Lm}$ are not present as in the previous case with $^6$Li. The parent states are denoted by $^{2T+1,2S+1}L_P$, which include $^{13}S$, $^{31}S$, $^{13}D$, and $^{31}D$ two-nucleon states. The products of two one-particle fractional parentage coefficients are replaced by the squares of each single coefficient because the initial and final states belong to the same doublet $^{22}P$. These coefficients were given in Refs. [2,10]. The transition matrix elements are calculated for the states $n_i = n_f = 1$, $L_i = L_f = 1$, $l_i = l_f = 1$, $S_i = S_f = 1/2$, $J_i = J_f = 3/2$, $T_i = T_f = 1/2$, and $M_{T_i} = M_{T_f} = \mp 1/2$, where the minus and plus signs in the isospin projection number are for $^7$Li and $^7$Be, respectively.

Analytical calculations indicate that all the multipole form factors depend on the nucleon form factors and parameters of the unified electroweak theory. The multipole form factors of $^7$Be are the same as those of $^7$Li but with the opposite signs for the isovector components. They can be written in the common form as

$$F_0^C(q) = \frac{1}{\sqrt{\pi}} \tilde{X}^F J_0, \quad (10a)$$

$$F_2^C(q) = \frac{3}{5\sqrt{\pi}} \tilde{X}^F J_2, \quad (10b)$$

$$F_1^M(q) = -\frac{\sqrt{5}q}{9\sqrt{2\pi}m_N}[\tilde{X}^F(J_0 + J_2) + 3\tilde{Y}^F(J_0 - \frac{3}{25}J_2)], \quad (10c)$$

$$F_3^M(q) = -\frac{3\sqrt{3}q}{5\sqrt{5\pi}m_N} \tilde{Y}^F J_2, \quad (10d)$$

$$A_1^C(q)=\frac{\sqrt{5}q}{3\sqrt{\pi}m_N}\tilde{Y}^A(J_0+\frac{6}{25}J_2), \quad (10e)$$

$$A_3^C(q)=\frac{9q}{5\sqrt{5\pi}m_N}\tilde{Y}^A J_2, \quad (10f)$$

$$A_1^E(q)=\frac{\sqrt{10}}{3\sqrt{\pi}}\tilde{Y}^A(J_0-\frac{3}{25}J_2), \quad (10g)$$

$$A_3^E(q)=\frac{6\sqrt{3}}{5\sqrt{5\pi}}\tilde{Y}^A J_2. \quad (10h)$$

Here we introduce the notations $\tilde{X}^F=3G_E^s\mp G_E^v$, $\tilde{Y}^F=G_M^s\pm G_M^v$ and $\tilde{Y}^A=\beta_A^{(0)}G_A^s\pm\beta_A^{(1)}G_A^v$ with the upper sign for $^7$Li and the lower sign for $^7$Be. The radial integrals $J_0(q)$ and $J_2(q)$ have the same expressions mentioned above when using the harmonic oscillator wavefunctions. The vector form factors can be derived from Eqs. (10a)-(10d) by replacing $\tilde{X}^F\rightarrow\tilde{X}^V=3\beta_V^{(0)}G_E^s\mp\beta_V^{(1)}G_E^v$ and $\tilde{Y}^F\rightarrow\tilde{Y}^V=\beta_V^{(0)}G_M^s\pm\beta_V^{(1)}G_M^v$. All the form factors have two isospin components, so the axial form factors are non-zero when applying the Weinberg-Salam model.

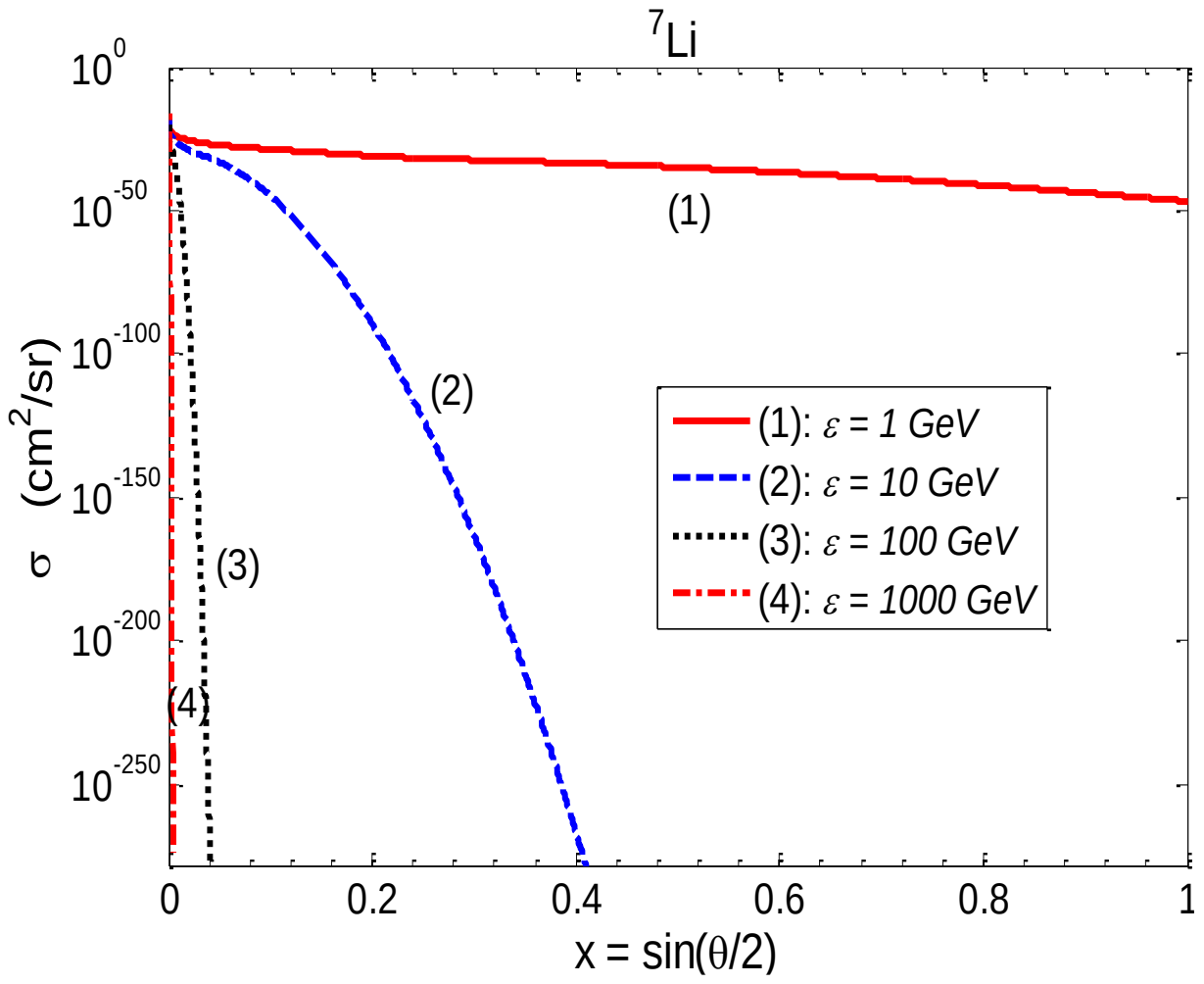

Fig. 2. Electron elastic scattering cross sections for $^7$Li obtained within the unified electroweak theory at 1, 10, 100, and 1000 GeV and different scattering angles.

The calculated values of the elastic scattering cross sections for $^7$Li and $^8$Be are given in Tables 1 and 2. The parameter $\beta$ = 1.3055 fm$^{-2}$ is selected for both cases following the same procedure described in Sec. 3.1, namely $\beta$ is adjusted so that the theoretical (elastic + quasi-elastic) cross section $\sigma_{\rm theor}$ = 2.7907 x 10$^{-31}$ cm$^2$/sr at 600 MeV and 32$^0$ (see Table 1) agrees with the experimental one $\sigma_{\rm exp}$ = (2.79 ± 0.1) x 10$^{-31}$ cm$^2$/sr given in Table II of Ref. [5]. In general, the scattering cross sections of $^7$Li and $^7$Be vary following a similar rule, and both of them have slightly larger magnitudes than those of $^6$Li. The difference becomes considerable only at the large energies and scattering angles. Here, we assume that the nuclei with the same number of nucleons have the harmonic oscillation potentials with the same frequency.

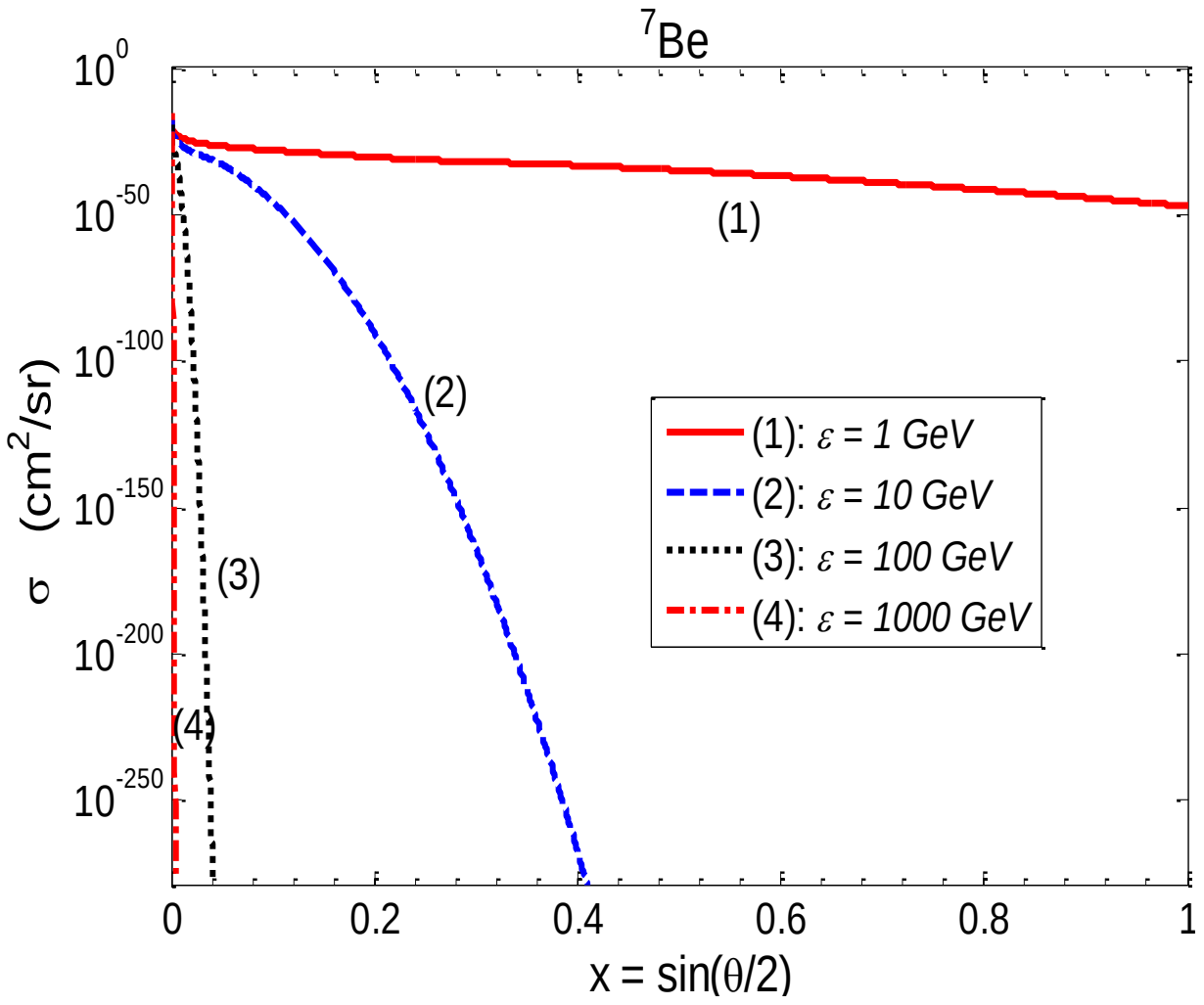

Fig. 3. Electron elastic scattering cross sections for $^7$Be obtained within the unified electroweak theory at 1, 10, 100, and 1000 GeV and different scattering angles.

Figs. 2 and 3 show the elastic cross sections at 1, 10, 100, and 1000 GeV obtained for two nuclei under consideration. The extreme points are not found in the graphs above and in Tables 1 and 2. Although both share a similar fashion, the cross sections of $^7$Be are in practice slightly larger than those of $^7$Li, except at too large angles. It can be seen that there are only forward scattering processes when the incident electron energies are hundreds of GeV and higher. It should be noted that the scattering cross sections of $^7$Li obtained in the present study are moderately different from those in Ref. [20] due to the adjustment of the center of mass motion and the value of the selected harmonic oscillator parameter. Specifically, the present cross sections decrease with increasing the scattering angles faster than those in Ref. [20], particularly at the energy above 10 GeV. However, the transformation rule of both calculated cross sections remains nearly unchanged.

### 3.3. Quasi-elastic electron scattering on $^7$Li via the $3/2^- \rightarrow 1/2^-$ transition

We further investigate the quasi-elastic scattering of $^7$Li via the excitation from the $3/2^-$

ground state to the nearest $1/2^-$ excited state. The electromagnetic scattering of unoriented nuclei by unpolarized electrons at an energy of 197 MeV was also studied in detail for this transition in Ref. [2]. According to the selection rule, the order of the multipoles in the scattering under consideration must satisfy condition $1 \le L \le 2$ and there is no forbidden multipole. The numerical calculations are proceeded in the same manner as those in the elastic scattering but with $J_i = 3/2$ and $J_f = 1/2$. Primarily, the electromagnetic and vector form factors have expressions similar to those in Ref. [20], but the axial ones have the opposite sign.

Table 1. Electron scattering cross sections for $^6$Li, $^7$Li, $^7$Be, and $^7$Li$^*$ obtained within the unified electroweak theory at different energies (below 1 GeV) and scattering angles.

| $\sigma = \sigma_F + \sigma_{FZ} + \sigma_Z$ (cm$^2$/sr) | | | | | | | |
|---|---|---|---|---|---|---|---|
| $\varepsilon$ | Nuclei | $\theta = 0^0$ | $\theta = 32^0$ | $\theta = 60^0$ | $\theta = 90^0$ | $\theta = 120^0$ | $\theta = 180^0$ |
| 200 MeV | $^6$Li | 1.2976 x $10^{-15}$ | 1.556 x $10^{-29}$ | 5.9446 x $10^{-31}$ | 3.6578 x $10^{-32}$ | 2.9564 x $10^{-33}$ | 9.4169 x $10^{-35}$ |
| | $^7$Li | 1.2976 x $10^{-15}$ | 1.6025 x $10^{-29}$ | 6.8096 x $10^{-31}$ | 5.7481 x $10^{-32}$ | 9.6553 x $10^{-33}$ | 2.6295 x $10^{-33}$ |
| | $^7$Be | 5.1903 x $10^{-15}$ | 6.4016 x $10^{-29}$ | 2.6293 x $10^{-30}$ | 1.8468 x $10^{-31}$ | 1.7942 x $10^{-32}$ | 5.7003 x $10^{-34}$ |
| | $^7$Li$^*$ | 1.8005 x $10^{-24}$ | 2.0927 x $10^{-31}$ | 4.5335 x $10^{-32}$ | 1.2848 x $10^{-32}$ | 4.6214 x $10^{-33}$ | 1.8671 x $10^{-33}$ |
| 400 MeV | $^6$Li | 3.2439 x $10^{-16}$ | 1.6140 x $10^{-30}$ | 5.4472 x $10^{-33}$ | 2.0345 x $10^{-35}$ | 1.2200 x $10^{-35}$ | 2.0315 x $10^{-36}$ |
| | $^7$Li | 3.2439 x $10^{-16}$ | 1.8479 x $10^{-30}$ | 1.6446 x $10^{-32}$ | 1.1061 x $10^{-33}$ | 3.0080 x $10^{-34}$ | 8.1290 x $10^{-35}$ |
| | $^7$Be | 1.2976 x $10^{-15}$ | 7.3357 x $10^{-30}$ | 5.3307 x $10^{-32}$ | 1.8021 x $10^{-33}$ | 2.8162 x $10^{-34}$ | 4.5209 x $10^{-35}$ |
| | $^7$Li$^*$ | 1.8005 x $10^{-24}$ | 1.1882 x $10^{-31}$ | 7.7484 x $10^{-33}$ | 7.4773 x $10^{-34}$ | 1.6433 x $10^{-34}$ | 4.1906 x $10^{-35}$ |
| 600 Mev | $^6$Li | 1.4418 x $10^{-16}$ | 1.4815 x $10^{-31}$ | 9.3728 x $10^{-35}$ | 3.8287 x $10^{-36}$ | 7.0513 x $10^{-38}$ | 1.7747 x $10^{-39}$ |
| | $^7$Li | 1.4418 x $10^{-16}$ | 2.2936 x $10^{-31}$ | 1.6065 x $10^{-33}$ | 6.9194 x $10^{-35}$ | 2.7019 x $10^{-36}$ | 9.9981 x $10^{-38}$ |
| | $^7$Be | 5.7670 x $10^{-16}$ | 8.8764 x $10^{-31}$ | 3.4757 x $10^{-33}$ | 7.4337 x $10^{-35}$ | 1.7017 x $10^{-36}$ | 4.8537 x $10^{-38}$ |
| | $^7$Li$^*$ | 1.8005 x $10^{-24}$ | 4.9709 x $10^{-32}$ | 1.0582 x $10^{-33}$ | 3.5254 x $10^{-35}$ | 1.3446 x $10^{-36}$ | 4.9404 x $10^{-38}$ |
| 1 GeV | $^6$Li | 5.1903 x $10^{-17}$ | 1.4153 x $10^{-34}$ | 5.3075 x $10^{-37}$ | 1.3271 x $10^{-41}$ | 7.1254 x $10^{-46}$ | 3.3982 x $10^{-50}$ |
| | $^7$Li | 5.1903 x $10^{-17}$ | 7.4870 x $10^{-33}$ | 1.2000 x $10^{-35}$ | 1.2621 x $10^{-39}$ | 1.3924 x $10^{-43}$ | 1.5494 x $10^{-47}$ |
| | $^7$Be | 2.0761 x $10^{-16}$ | 2.1114 x $10^{-32}$ | 1.3370 x $10^{-35}$ | 6.1272 x $10^{-40}$ | 5.7247 x $10^{-44}$ | 7.0048 x $10^{-48}$ |
| | $^7$Li$^*$ | 1.8005 x $10^{-24}$ | 5.6524 x $10^{-33}$ | 5.9891 x $10^{-36}$ | 6.0050 x $10^{-40}$ | 6.5536 x $10^{-44}$ | 7.5919 x $10^{-48}$ |

As shown in Tables 1 and 2, the quasi-elastic scattering cross section varies in the same way as the elastic scattering one but with smaller magnitudes. The difference between both cases is only considerable when the scattering occurs at extremely small angles, especially at $0^0$. Meanwhile, there are similarities in the orders of magnitude of the cross sections for the remaining angles. We note here that the experiment was able to measure the total (quasi-elastic + elastic) scattering cross sections only, while our theoretical calculation is able to distinguish them, and thus can analyze their separate contributions to the total cross section data. In general, the scattering cross sections of four cases decrease rapidly with increasing the energy and scattering angle, especially at high energies. This is because the used nucleon form factors have been parameterized in terms of dipoles, which depend on the square of the momentum transfer. MATLAB source files of all cross sections are given in the Supplemental Material, Ref. [31].

Table 2. Electron scattering cross sections for $^6$Li, $^7$Li, $^7$Be, and $^7$Li$^*$ obtained within the unified electroweak theory at high energies (10, 100, and 1000 GeV) and small scattering angles.

| $\sigma = \sigma_F + \sigma_{FZ} + \sigma_Z$ (cm$^2$/sr) | | | | | |
|---|---|---|---|---|---|
| $\varepsilon$ | Nuclei | $^6$Li | $^7$Li | $^7$Be | $^7$Li$^*$ |
| 10 GeV | $\theta = 0^0$ | 5.1898 x $10^{-19}$ | 5.1899 x $10^{-19}$ | 2.0759 x $10^{-18}$ | 1.8004 x $10^{-24}$ |
| | $\theta = 2^0$ | 2.4835 x $10^{-29}$ | 4.1292 x $10^{-29}$ | 1.6345 x $10^{-28}$ | 1.0386 x $10^{-29}$ |
| | $\theta = 3^0$ | 6.3607 x $10^{-34}$ | 1.0666 x $10^{-30}$ | 3.3791 x $10^{-30}$ | 8.4202 x $10^{-31}$ |
| | $\theta = 32^0$ | 2.1112 x $10^{-155}$ | 1.8191 x $10^{-143}$ | 3.9986 x $10^{-144}$ | 5.1377 x $10^{-144}$ |
| 100 GeV | $\theta = 0^0$ | 5.1422 x $10^{-21}$ | 5.1464 x $10^{-21}$ | 2.0587 x $10^{-20}$ | 1.7894 x $10^{-24}$ |
| | $\theta = 2^0$ | 1.4419 x $10^{-79}$ | 2.7431 x $10^{-74}$ | 5.0209 x $10^{-75}$ | 8.3314 x $10^{-75}$ |
| | $\theta = 3^0$ | 2.0478 x $10^{-141}$ | 1.4273 x $10^{-130}$ | 3.0663 x $10^{-131}$ | 4.0290 x $10^{-131}$ |
| 1000 GeV | $\theta = 0^0$ | 2.0085 x $10^{-23}$ | 2.2120 x $10^{-23}$ | 8.8852 x $10^{-23}$ | 9.8782 x $10^{-25}$ |
| | $\theta = 1^0$ | 0 | 0 | 0 | 0 |

We show the scattering cross sections of three nuclei at 10 GeV and higher in a separate Table 2. As can be easily seen, at this energy scale, the obtained cross sections quickly approach zero values at large scattering angles, namely above $32^0$ at 10 GeV, $4^0$ at 100 GeV, and $1^0$ at 1000 GeV. The scattering angle of $32^0$ is highlighted because the selected oscillator parameters and the calculated cross sections show the best descriptions for the experimental data in Ref. [5].

## IV. DISCUSSIONS

By comparing the calculated elastic scattering cross sections for different light nuclei, we can see that their values generally increase in the orders of $^6$Li, $^7$Li, and $^7$Be at all energies and scattering angles. On the aspect relating to the geometric size of the object, that the scattering cross sections of $^7$Li and $^7$Be are larger than those of $^6$Li is consistent with the rule that the nuclear size is proportional to the mass number, namely $R \sim A^{1/3}$. Moreover, the scattering cross sections of $^7$Be are slightly larger than those of $^7$Li even though both have the same mass number. This is because neutrons behave like neutral particles and the electromagnetic cross section is mainly contributed by the charge distribution of protons. In addition, our results also indicate that the detectable size of a target nucleus is unfixed, but depends on the energies of the projectile. It will go down as the energies go up, implying that electrons can penetrate inside the nuclei and nucleons, and tend to go straight when being scattered at high energies. On the aspect of probability, this statement is further reinforced since the scattering cross sections at the energies above tens of GeV only differ from zero at very small angles and are almost equal to zero at higher ones.

Although the weak interactions have an insignificant contribution to the scattering cross section at MeV energies, the present work simultaneously treats the electromagnetic and weak interactions into one expression, which can be investigated for arbitrary energies. In particular, no approximation has been employed and we do not ignore the convection currents when performing the calculations for all investigated cases. In addition, when the axial form factors are computed in the same way as in Ref. [13], they lead to the results which differ from those obtained above by a factor $q/2m_N$. This shows the difference between our directly calculated results and those obtained by using the conventional impulse approximation. However, when the momentum transfer equals to twice the nucleon mass, the axial form factors and scattering cross sections computed by using both methods give the same results. This proves that the impulse approximation can be applied to the electron-nucleus scattering when the condition $q \leq 2m_N$ is satisfied since the contribution of the weak interactions remains negligible. The scattering cross sections obtained in two manners start to deviate if $q > 2m_N$, especially at high energies and large angles due to $q = 2\varepsilon x$.

As reported in Ref. [5], the experimental cross sections determined at various energies and angles have different accuracies. For one nucleus, fitting the calculated cross sections at a fixed angle to the experimental data at various energies should lead to different values of the harmonic oscillator parameter. The analyses in Refs. [2,5] also show that the well-described oscillator parameter for light nuclei is not fixed, but varies within a limited range. In addition, matching the calculated cross sections to the experimental data of $^6$Li and $^7$Li at the same energy and scattering angle should also give different oscillator parameters for the two nuclei. These discussions suggest that the self-consistent potential field should fluctuate with various oscillator frequencies, depending on the incident electron energies, scattering angles, and certain target nuclei. In other words, the oscillator parameter in practice depends on the momentum transfer and nuclear mass number. Thus, finding an appropriate expression that can express their relation would allow us to describe the experimental data better.

## V. SUMMARY AND OUTLOOK

We have already presented a complete calculation method for the electron-nucleus scattering cross section within the unified electroweak theory, which can be applied to all light nuclei. The many-particle shell model and fractional parentage coefficients are used to directly calculate the multipole form factors without using any approximations. With an appropriate adjustment for the harmonic oscillator parameter, the obtained results can give a good description of the experimental data. In addition, the rule that the nuclear size increases with increasing the mass number is also confirmed by comparing the elastic cross sections of investigated nuclei at a certain energy and scattering angle. In particular, theoretical calculations also predict that, with energies above hundreds of GeV, electrons can penetrate through nucleons and go straightforward rather than knock them out of (break down) nuclei as often described in the scattering processes at low energies.

This study can be further extended by considering more transitions in the same nucleus and/or other light nuclei, especially when the orientation of the nuclear spin and the charge-changing weak currents are taken into account. Such an extension helps elucidate the role and the contribution of each constituent (neutral, charge-changing weak, and electromagnetic currents) to the beta decay processes inside unstable nuclei. In addition, once our predictions are experimentally confirmed, new prospects will be opened in studying the structure of nuclei by high-energy electrons using the unified electroweak theory and studying the weak interactions at the nuclear scale. Calculating the multipole form factors in electron-nucleus scattering along with other relevant quantities, which utilizes one or various theoretical models of the interaction unification, would be an extensive realm in theoretically nuclear physics.

## ACKNOWLEDGMENTS

One of the authors (M.T. Vo) expresses his gratitude to Dr. Sc. Luong Zuyen Phu (Z.P. Luong) for his guidance and suggestions on this work.

## REFERENCES

[1]. R. Hofstadter, Annu. Rev. Nucl. Sci. **7**, 231 (1957).

[2]. R.S. Willey, Nucl. Phys. **40**, 529 (1963).

[3]. L.J. Weigert, M.E. Rose, Nucl. Phys. **51**, 529 (1964).

[4]. R.E. Rand, R. Frosch, M.R. Yearian, Phys. Rev. **144**, 859 (1966).
[5]. L.R. Suelzle, M.R. Qearian, H. Crannell, Phys. Rev. **162**, 992 (1967).
[6]. N.N. Bogoliubov, D.V. Shirkov, *Quantum Fields*, Benjamin/Cummings Publishing Company Inc. USA 1983.
[7]. J.C. Gaborit, M. Silari, L. Ulrici, Nucl. Instrum. Meth. Phys. Res. A **565**, 333 (2006).
[8]. J.T. Seeman, Annu. Rev. Nucl. Part. Sci. **41**, 389 (1991).
[9]. A.I. Akhiezer, A.G. Sitenko, V.K. Tartakovsky, *Nuclear Electrodynamics*, Springer Berlin 1994.
[10]. H.A. Jahn, H. van Wieringen, Proc. R. Soc. A: Math. Phys. Sci. **209**, 502 (1951).
[11]. J.P. Elliott, J. Hope, H.A. Jahn, Philos. Trans. R. Soc. A **246**, 241 (1953).
[12]. P.J. Redmond, Proc. R. Soc. A: Math. Phys. Sci. **222**, 84 (1954).
[13]. B.K. Kerimov, A.Z. Agalarov, M.Y. Safin, Sov. Phys. J. **27**, 327 (1984).
[14]. Z.P. Luong, Nucl. Phys. A **722**, 419c (2003).
[15]. Z.P. Luong, Bull. Russ. Acad. Sci.: Phys. Ser. **67**, 1495 (2003).
[16]. T.W. Donnelly, J.D. Walecka, Phys. Lett. B **41**, 275 (1972).
[17]. T.W. Donnelly, J.D. Walecka, Phys. Lett. B **44**, 330 (1973).
[18]. T.W. Donnelly, R.D. Peccei, Phys. Rep. **50**, 1 (1979).
[19]. C.J. Horowitz, D.P. Murdock, B.D. Serot, *The Relativistic Impulse Approximation*. In: K. Langanke, J.A. Maruhn, S.E. Koonin, (eds) *Computational Nuclear Physics* 1, Springer Berlin 1991.
[20]. Z.P. Luong, M.T. Vo, Phys. Lett. B **844**, 138095 (2023).
[21]. T.W. Donnelly, A.S. Raskin, Ann. Phys. (USA) **169**, 247 (1986).
[22]. O. Benhar, V.R. Pandharipande, Phys. Rev. C **47**, 2218 (1993).
[23]. Z.-G. Wang, S.-L. Wan, W.-M. Yang, Eur. Phys. J. C **47**, 375 (2006).
[24]. B. Povh, K. Rith, C. Scholz, F. Zetsche, *Particles and Nuclei*, Springer Berlin 2008.
[25]. E. Chang, Z. Davoudi, W. Detmold, A.S. Gambhir, K. Orginos, M.J. Savage, P.E. Shanahan, M.L. Wagman, F. Winter, Phys. Rev. Lett. **120**, 152002 (2018).
[26]. M.T. Burkey, *Searching for Tensor Currents in the Weak Interaction Using Lithium-8 β Decay*, Chicago Illinois 2019.
[27]. O. Tomalak, Phys. Rev. D **103**, 013006 (2021).
[28]. F. He, I. Zahed, Phys. Rev. C **109**, 045209 (2023).
[29]. R. Wang, C. Han, X. Chen, Phys. Rev. C **109**, L012201 (2024).
[30]. L.J. Tassie, F.C. Barker, Phys. Rev. **111**, 940 (1958)
[31]. See Supplemental Material at [URL will be inserted by publisher] for Matlab source files of the scattering cross sections for $^{6,7}$Li and $^{7}$Be.